\documentstyle[twocolumn,seceq,epsbox]{jpsj}

\def\runtitle
{
  Quantum Monte Carlo Simulation of the Trellis Lattice\ldots
  }
\def\runauthor{
  Shin {\sc Miyahara},
  Matthias {\sc Troyer},
  David C. {\sc Johnston}
  and Kazuo {\sc Ueda}
  }

\title
{
  Quantum Monte Carlo Simulation of the Trellis Lattice Heisenberg Model for SrCu$_2$O$_3$ and 
  CaV$_2$O$_5$
  }

\author
{
  Shin {\sc Miyahara}, 
  Matthias {\sc Troyer},
  David C. {\sc Johnston}$^{1}$
  and Kazuo {\sc Ueda}
}

\inst
{
  Institute for Solid State Physics, University of Tokyo,\\
  Roppongi 7-22-1, Minatoku, Tokyo 106-8666, Japan \\
  $^1$Ames Laboratory-USDOE and Department of Physics and Astronomy \\
  Iowa State University, Ames, Iowa 50011, U.S.A.
}

\recdate
{
  \today
}

\abst { We study the spin-$1/2$ trellis lattice Heisenberg model, a
  coupled spin ladder system, both by perturbation around the dimer
  limit and by quantum Monte Carlo simulations. We discuss the
  influence of the inter-ladder coupling on the spin gap and the
  dispersion, and present results for the temperature dependence of the
  uniform susceptibility. The latter was found to be parameterized
  well by a mean-field type scaling ansatz. Finally we discuss fits of
  experimental measurements on SrCu$_2$O$_3$ and CaV$_2$O$_5$ to our
  results.  }

\kword
{
  spin gap, spin ladder, trellis lattice, Heisenberg antiferromagnet,
  SrCu$_2$O$_3$, CaV$_2$O$_5$, quantum Monte Carlo, scaling}

\begin{document}
\maketitle

\section{Introduction}
Spin-$1/2$ ladder models\cite{Daggot-Rice} can describe the magnetic
behavior of a variety of quasi-one dimensional materials. Examples
include the cuprate materials SrCu$_2$O$_3$~\cite{M.Azuma} and
LaCuO$_{2.5}$ \cite{Nature} and the vanadate
CaV$_2$O$_5$.\cite{H.Iwase} Spin excitations in isolated ladders have
a finite energy gap, which makes them prototypical spin
liquids.\cite{Daggot-Rice} This is of interest in relation with high
temperature superconductors, since upon doping they become doped
resonating-valence-bond liquids, with a spin excitation gap and
dominating quasi-long range pairing correlations.

While isolated ladders are relatively easy to study and offer a
variety of interesting phenomena, it is still necessary to study the
inter-ladder coupling. Although the inter-ladder coupling is weak it
is still necessary to obtain superconducting long range order, as
observed in the ladder compound (Sr,Ca)$_{14}$Cu$_{24}$O$_{41}$
\cite{142441}.  Even in undoped ladder systems the weak inter-ladder
coupling can play an important role. In LaCuO$_{2.5}$ an inter-ladder
coupling of only one tenth of the intra-ladder coupling is sufficient
to destroy the spin gap of the isolated ladder and leads to
antiferromagnetic long range order.\cite{lacuo}

One of the authors fitted the uniform magnetic susceptibility of
SrCu$_2$O$_3$ to that of an isolated ladder model\cite{Johnston} and
found that
the best fit is achieved with a ratio $J_{\bot}/J_{\|} \approx
0.5$, where $J_{\bot}$ is the coupling along a rung of the ladder and
$J_{\|}$ that along the legs of the ladder.
This result is surprising since the Cu-Cu distance
across a rung is shorter than that along a leg and
the rung coupling $J_{\bot}$ thus
expected to be larger than the leg coupling.
It was suggested
that this apparent ratio of $J_{\bot}/J_{\|} \approx 0.5$ may be due
to the neglect of inter-ladder interactions, although a mean-field
analysis of the influence of these interactions suggested a similar
ratio.
In this paper we report on an
investigation of the effect of the inter-ladder coupling on the
magnetic susceptibility and the magnon dispersion of coupled ladders
and dimers. Fits of the results to experimental measurements reveal
that the inter-ladder coupling does not change the original estimates
significantly and will be discussed in detail in a forthcoming
publication.\cite{Johnston2}

The susceptibility of another ladder-like compound 
CaV$_2$O$_5$ was fitted by Onoda and Nishiguchi to an isolated
dimer model with $J_{\|}=0$.\cite{Onoda} Again
this result is surprising as the bond lengths are similar. We perform
simulations also for this compound to estimate the strength of
inter-dimer couplings.
\begin{figure}
  \begin{center}
    \psbox{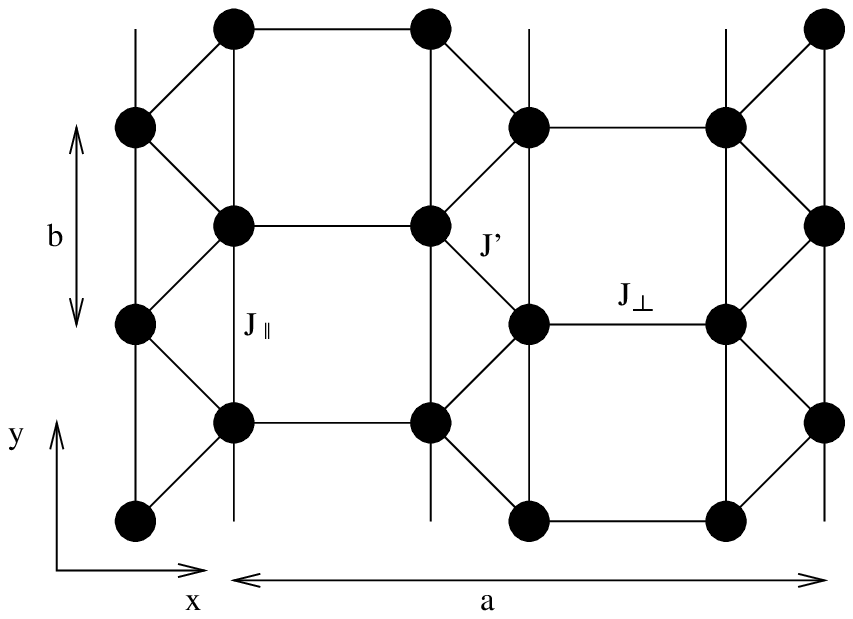}
  \end{center}
    \caption{The lattice structure of the trellis lattice
      Heisenberg model. $J_{\bot}$ is the exchange constant
      across the rung, $J_{\|}$ is that along the legs and
      $J'$ is that between ladders.}
  \label{fig:model}
\end{figure}

Figure~\ref{fig:model} shows the structure of the trellis lattice.  It
consists of spin ladders (with coupling along the legs $J_{\|}$ and
across the rungs $J_{\bot}$, coupled by frustrating inter-ladder
``zig-zag'' couplings $J'$. The size of the lattice used in
the calculations is denoted by
$N_x \times N_y$, where $N_x$ is the number of the spin along the
x-axis and $N_y$ along the y-axis.

In SrCu$_2$O$_3$ and CaV$_2$O$_5$ the interactions in the ladder ($J_{\bot}$ 
and $J_{\|}$) are antiferromagnetic. The inter-ladder coupling ($J^{\prime}$)
in SrCu$_2$O$_3$ is expected to be ferromagnetic, since it occurs via $90^{\circ}$
Cu-O-Cu bonds. In CaV$_2$O$_5$ the local chemistry is more complex, 
and the sign of the exchange coupling $J'$ has not been definitively 
established.

\section{Spin Gap and Magnon Dispersion}

In this section, we discuss the influence of the inter-ladder coupling 
$J^{\prime}$ on the spin gap and the magnon dispersion.  Following 
refs.  \cite{barnes92,reigrotzki}, but extending their ideas to include the 
inter-ladder couplings, we start from the coupled dimer system, where 
$J_{\bot} \gg$ $J_{\|}$, $J^{\prime}$.

The ground state in the dimer limit ($J_{\|}=J^{\prime}=0$) consists
of spin singlets on the rungs, with an energy per rung of $E_{singlet}
= - \frac{3}{4} J_{\bot}$.  Including the inter-dimer couplings
perturbatively, the ground state energy per rung to second order is
\begin{equation}
  E_0 = -\frac{3}{8} J_{\bot} - \frac{3}{16} \frac{J_{\|}^2}{J_{\bot}}
  - \frac{3}{32} \frac{{J^{\prime}}^2}{J_{\bot}}.
\end{equation}

The lowest lying excitations are spin-$1$ magnons, where one of the
rung singlets is turned into a triplet. These magnons disperse due to
inter-dimer couplings. The lowest branch is
\begin{eqnarray}
  \omega (k_{\bot}, k_{\|}) & = &  J_{\bot} 
                  + \frac{3}{4}\frac{{J_{\|}}^2}{J_{\bot}}
                  - \frac{1}{8}\frac{{J^{\prime}}^2}{J_{\bot}}
                  \nonumber \\
            &   & + J_{\|} \cos(k_{\|})
                  - \frac{1}{4}\frac{{J_{\|}}^2}{J_{\bot}} \cos(2k_{\|})
                  \nonumber \\
            &   & + (-|J^{\prime}| - \frac{|J^{\prime}|J^{\prime}}{2J_{\bot}} 
                  + \frac{J_{\|}|J^{\prime}|}{2 J_{\bot}})
                  \cos(\frac{k_{\bot}}{2}) \cos(\frac{k_{\|}}{2})
                  \nonumber \\
            &   & -\frac{{J^{\prime}}^2}{8J_{\bot}}
                  ( \cos(k_{\bot})\cos(k_{\|})+\cos(k_{\bot})+\cos(k_{\|}) )
                  \nonumber \\
            &   & + \frac{J_{\|}|J^{\prime}|}{2 J_{\bot}}
                  \cos(\frac{k_{\bot}}{2}) \cos(\frac{3k_{\|}}{2}),
                  \label{eq:spingap}
\end{eqnarray}
where we have chosen units such that the lattice parameters (see
Fig.~\ref{fig:model}) are $a = b = 1$.

Minimizing $\omega (k_{\bot}, k_{\|})$, we obtain for the the spin gap
$\Delta_s$ in next to leading order
\begin{equation}
  \Delta_s = \left\{
  \begin{array}{@{\,}ll}
    J_{\bot} - J_{\|} - \frac{{J^{\prime}}^2}{8 J_{\|}}
      & J_{\|} \geq | 0.25 J^{\prime} | \\
    J_{\bot} + J_{\|} - |J^{\prime}| & J_{\|} \leq | 0.25 J^{\prime} |.
  \end{array}
  \right. 
  \label{eq:spingap_first}
\end{equation}
The minimum of the dispersion is at momenta
\begin{eqnarray}
  k_{\bot} &=& 0 \\
  k_{\|} &=& \left\{
  \begin{array}{ll}
    2 \arccos{\frac{J^{\prime}}{4 J_{\|}}}
    & J_{\|} \geq | 0.25 J^{\prime} | \\
    0 & J_{\|} \leq | 0.25 J^{\prime} |. \\
  \end{array}
  \right.
\end{eqnarray}

We can extend above results to arbitrary ratios of 
$J_{\|}/J_{\bot}$ by replacing the second order 
expression of the dispersion due to $J_{\|}$
\begin{eqnarray}
  \epsilon^{(2)} (k_{\|}) & = & J_{\bot}
  + \frac{3}{4}\frac{{J_{\|}}^2}{J_{\bot}} \nonumber \\
  &   & + J_{\|} \cos(k_{\|}) 
  - \frac{1}{4}\frac{{J_{\|}}^2}{J_{\bot}}
  \cos(2k_{\|})
\end{eqnarray}
by the exact dispersion $\epsilon (k_{\|})$.

Substituting the $J_{\|}$ terms in eq.~(\ref{eq:spingap}) by $\epsilon 
(k_{\|})$ we obtain
\begin{eqnarray}
   \tilde{\omega} (k_{\bot}, k_{\|}) & = & \epsilon (k_{\|}) 
                  - \frac{1}{8}\frac{{J^{\prime}}^2}{J_{\bot}}
                  \nonumber \\
            &   & + \left[-2 |J^{\prime}|
                  - \frac{|J^{\prime}|J^{\prime}}{2J_{\bot}}
                  + \frac{|J^{\prime}|}{J_{\bot}} \epsilon (k_{\|})\right] 
                  \nonumber \\
                  &&\qquad\times
                  \cos(\frac{k_{\bot}}{2}) \cos(\frac{k_{\|}}{2})
                  \nonumber \\
            &   & -\frac{{J^{\prime}}^2}{8J_{\bot}}
                  \left[\cos(k_{\bot})\cos(k_{\|})+\cos(k_{\bot})
                   \right.\nonumber \\
                   & & \left.\qquad\qquad +\cos(k_{\|})\right].
                  \label{eq:dispersion general}
\end{eqnarray}

Expanding $\epsilon(k_{\|})$ around the minimum at $k_{\|}=\pi$ as
\begin{equation}
\epsilon (\pi + k_{\|})
         = \sqrt{{\Delta_0}^2 + v^2 {k_{\|}}^2},
\end{equation}
where $\Delta_0$ is the spin gap for the spin ladder system and $v$ is 
the magnon velocity we get a minimum in the dispersion at
\begin{equation}
\pi + (2 - \frac{\Delta_0}{J_{\bot}})
\frac{{\Delta_0}^2}{v^2}
{(\frac{J^{\prime}}{\Delta_0})},
\end{equation}
and for the influence of the inter-ladder 
coupling $J'$ on the spin gap
\begin{equation}
  \Delta_s = \Delta_0 \{ 1 - \frac{1}{2}
  {(\frac{\Delta_0}{v})}^2 C (\frac{J^{\prime}}{\Delta_0})^2 \},
  \label{eq:spingapgeneral}
\end{equation}
where
\begin{equation}
  C = {(2 - \frac{\Delta_0}{J_{\bot}})}^{2}.
\end{equation}
The change in the spin gap is small, second order in $J^{\prime}$ and with
a typically small prefactor
$\frac{1}{2}{(\frac{\Delta_0}{v})}^2\approx0.1$.
~\cite{barnes92,Troyer}

We have calculated the spin gap on $(0,\pi)$ on $4 \times 4$ and $4
\times 6$ lattices by exact diagonalization using the Lanczos
algorithm\cite{Lanczos} and observed only minimal changes in the spin
gap, in agreement with above arguments.  These system sizes are
however too small to allow quantitative comparisons.

\section{The Uniform Susceptibility}

\subsection{Weakly Coupled Ladders}

While the inter-ladder coupling $J^{\prime}$ ($J^{\prime} \leq 0.2 J_{\|}$)
has negligible influence on the 
spin gap it still modifies the uniform magnetic susceptibility $\chi$
at intermediate and high temperatures. Using the quantum Monte Carlo 
loop algorithm\cite{Evertz,Evertz2} with improved estimators
\cite{Ammon} we calculated the temperature dependence of $\chi$.

We considered ladder systems with both $J_{\bot}/J_{\|} = 1$ and the
experimentally relevant coupling $J_{\bot}/J_{\|} =
0.5$.\cite{Johnston} The inter ladder couplings were chosen to be
$J^{\prime}/J_{\|}=\pm 0.1$, $\pm 0.2$, $\pm0.5$ for the isotropic ladders
and $J^{\prime}/J_{\|}=\pm 0.1$ and $\pm 0.2$ for $J_{\bot}/J_{\|} = 0.5$.

The quantum Monte Carlo simulations suffer from a negative sign
problem due to the frustrated inter-ladder couplings $J'$. This sign
problem was alleviated a bit by using improved estimators both for
$\chi$ and for the average sign.\cite{Ammon} Still the sign problem
restricts the QMC simulations to rather small lattices and not too low
temperatures.  In the temperature regime where QMC simulations were
feasible a lattice size of $4 \times 16$ was sufficiently large and
our results are not biased by finite size effects.
\begin{figure}
  \begin{center}
    \psbox{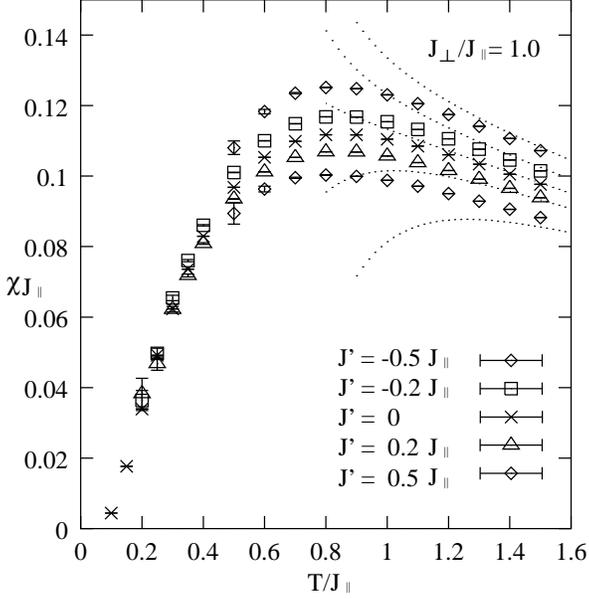}
  \end{center}
    \caption{Temperature dependence of the uniform susceptibility $\chi$
      of the trellis lattice Heisenberg model with $J_{\bot}/J_{\|} = 1$
      and $J'/J_{\|}=0,\pm 0.2$ and $\pm 0.5$.
      The dotted lines are fourth-order high temperature expansions.}
    \label{fig:susc_1.0}
\end{figure}
\begin{figure}
  \begin{center}
    \psbox{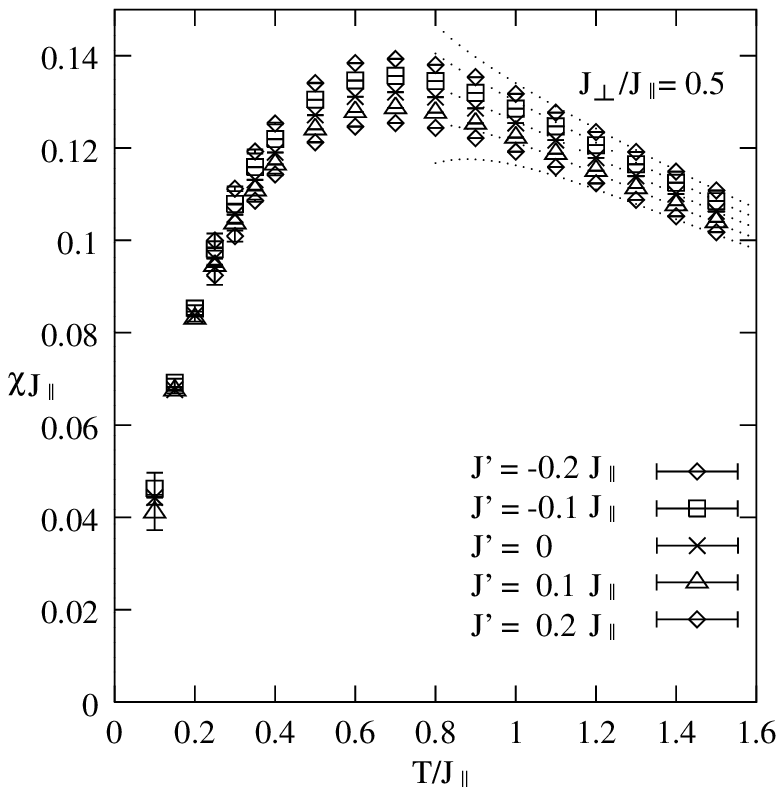}
  \end{center}
    \caption{Temperature dependence of the uniform susceptibility $\chi$
      of the trellis lattice Heisenberg model with $J_{\bot}/J_{\|} = 
      0.5$
      and $J'/J_{\|}=0,\pm 0.2$.
      The dotted lines are fourth-order high temperature expansions.}
    \label{fig:susc_0.5}
\end{figure}

Our results are shown in Figs.~\ref{fig:susc_1.0} and
~\ref{fig:susc_0.5} together with a fourth-order high temperature
expansion
\begin{eqnarray}
  \chi(T) & = & \frac{1}{4T} - \frac{1}{16} (J_{\bot} + 2 J_{\|} 
                + 2 J^{\prime}) (\frac{1}{T})^2 \nonumber \\
          &   & + \frac{1}{64} (4 J_{\bot} J_{\|} 
                + 4 J_{\bot} J^{\prime} + 8 J_{\|} J^{\prime} - {J_{\bot}}^2)
                (\frac{1}{T})^3 \nonumber \\
          &   & - \frac{1}{768} (- {J_{\bot}}^3 - 8 {J_{\|}}^3 
                - 8 {J^{\prime}}^3 - 6 {J_{\bot}}^2 {J_{\|}}  \nonumber \\
          &   & - 6 {J_{\bot}}^2 {J^{\prime}}
                + 24 {J_{\|}}^2 {J^{\prime}} + 12 {J_{\bot}} {J_{\|}}^2
                + 12 {J_{\bot}} {J^{\prime}}^2  \nonumber \\
          &   & + 6 J_{\|} {J^{\prime}}^2 
                + 72 J_{\bot} J_{\|} J^{\prime} )(\frac{1}{T})^4.
          \label{eq:highT}      
\end{eqnarray}

Fitting magnetic susceptibility measurements on spin 
ladder materials is much simplified if approximate analytic
expressions for the magnetic susceptibility are available.
The single ladder susceptibilities have been parameterized
before.\cite{Barnes2,Troyer,Johnston,Johnston2}

\begin{figure}
  \begin{center}
    \psbox{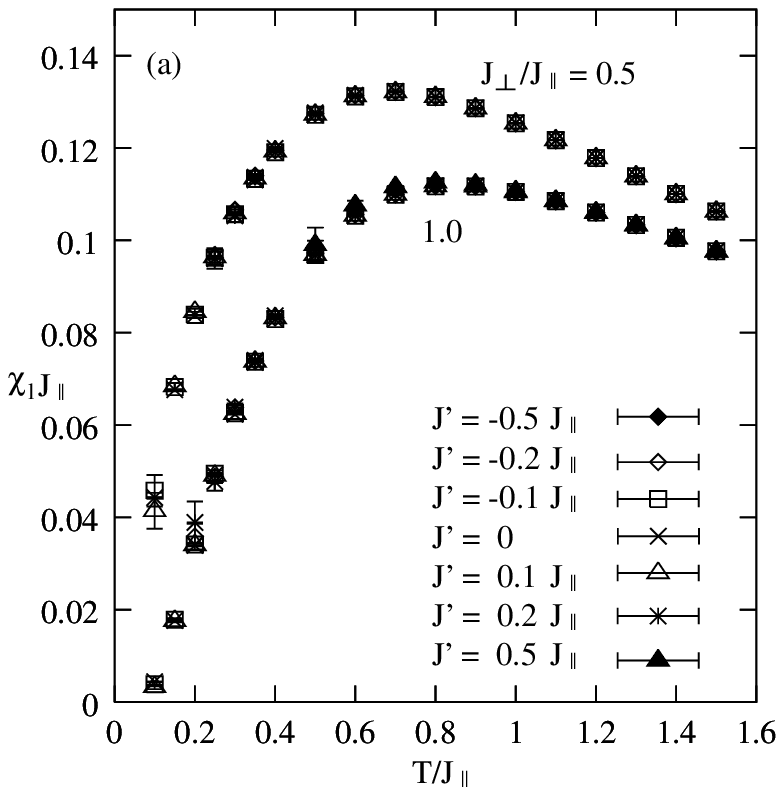}
    
    \psbox{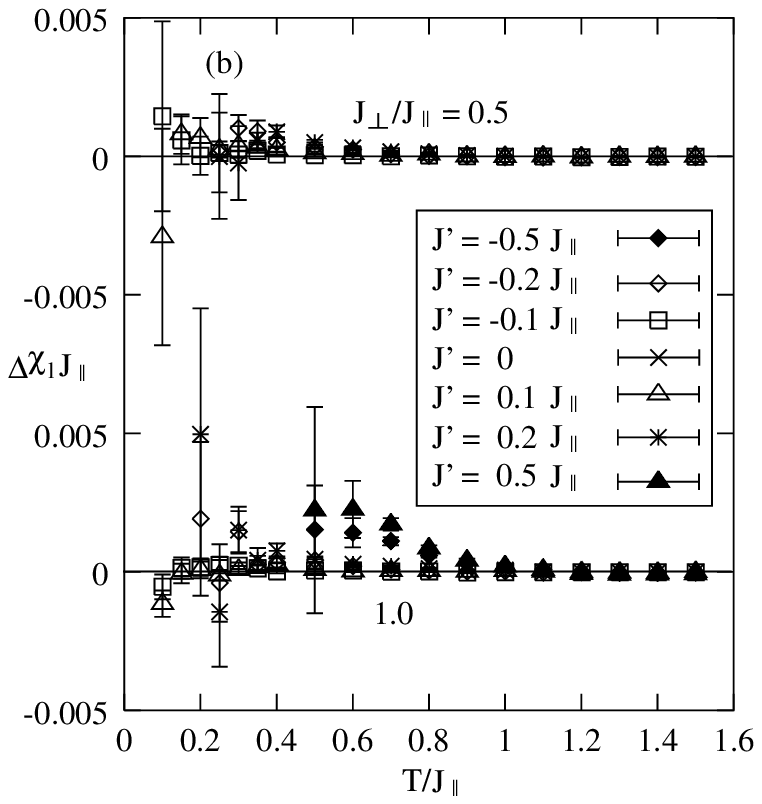}
  \end{center}
    \caption{(a) Scaling plot of weakly coupled ladders,
      (b) differences between the scaled and actual susceptibilities.}
    \label{fig:MFTScheck}
\end{figure}

To extend these parameterizations to include the inter-ladder coupling
we follow a mean field-type scaling ansatz (MFTS) of ref.
\cite{Johnston} and scale the trellis lattice susceptibility to
the susceptibility $\chi_l (J_{\bot}/J_{\|})$ of a single ladder:
\begin{equation}
  \chi(\frac{J_{\bot}}{J_{\|}}, \frac{J^{\prime}}{J_{\|}})
  = \frac{\chi_l(\frac{J_{\bot}}{J_{\|}})}
  {1+f(\frac{J^{\prime}}{J_{\|}}) \chi_l(\frac{J_{\bot}}{J_{\|}}) J_{\|}},
  \label{eq:MFTS}
\end{equation}
with
\begin{equation}
  f(\frac{J^{\prime}}{J_{\|}})
  = A \frac{J^{\prime}}{J_{\|}} + B (\frac{J^{\prime}}{J_{\|}})^2,
\end{equation}
and $A$ and $B$ as temperature independent constants. At low temperatures
this ansatz gives the same gapped behavior as a single ladder, which 
is reasonable since the gap is not changed much. Fixing $A=2$ also 
recovers the correct high temperature Curie-Weiss law. 
By fitting our QMC results in the temperature range
$1.0 \leq T/J_{\bot} \leq 1.5$ to eq.~(\ref{eq:MFTS}) we obtain
\begin{equation}
B = 0.3436(3).
\end{equation}
As a check of the quality of the MFTS ansatz we compare, in
Fig.~\ref{fig:MFTScheck}, the single ladder susceptibilities
obtained
from the QMC data of coupled ladders by inverting eq.~(\ref{eq:MFTS}) .
We find that the ansatz is useful for the whole temperature range as 
long as the inter-ladder coupling $J'$ is small.

\subsection{Coupled Dimers}

\begin{figure}
  \begin{center}
    \psbox{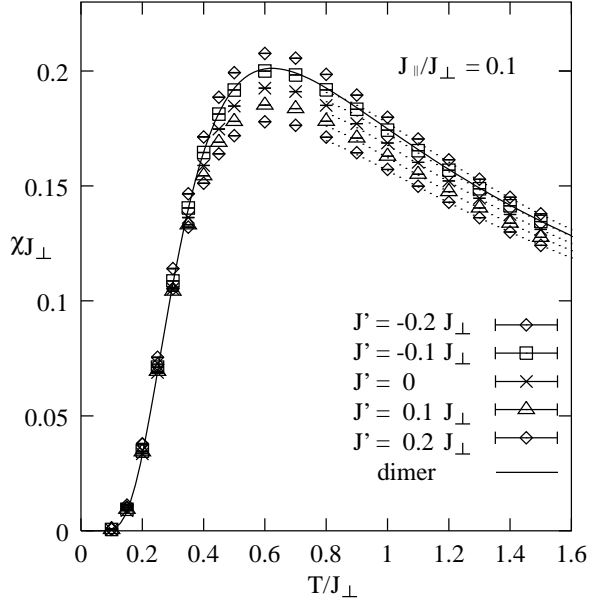}
  \end{center}
    \caption{Temperature dependence of the magnetic susceptibility
      of the trellis lattice Heisenberg  model in the regime of
      weakly coupled dimer. The coupling ratios are $J_{\|}/J_{\bot} = 0.1$
      and $J^{\prime}/J_{\bot}=\pm 0.1$ and $\pm 0.2$.
      The dotted lines are again the fourth order high temperature expansions.}
    \label{fig:susc_dimer}
\end{figure}

We can repeat a similar analysis for weakly coupled dimers instead of 
weakly coupled ladders.  We considered leg couplings of 
$J_{\|}/J_{\bot}=0$, $0.1$ and $0.2$ and inter-ladder couplings 
$J^{\prime}/J_{\bot}=0$, $\pm 0.1$ and $\pm 0.2$ on systems of size 
$32 \times 16$, where finite size effects are again negligible.  In 
Fig.~\ref{fig:susc_dimer} we show the uniform susceptibility for 
$J_{\|}/J_{\bot} = 0.1$.

We again make use of a MFTS ansatz to obtain an approximate analytic 
expression for the susceptibility
\begin{equation}
  \chi(\frac{J_{\|}}{J_{\bot}}, \frac{J^{\prime}}{J_{\bot}})
  = \frac{\chi_l(\frac{J_{\|}}{J_{\bot}})}
  {1+\tilde{f}(\frac{J^{\prime}}{J_{\bot}})
    \chi_l(\frac{J_{\|}}{J_{\bot}}) J_{\bot}},
  \label{eq:MFTS_dimer}
\end{equation}
with
\begin{equation}
  \tilde{f}(\frac{J^{\prime}}{J_{\bot}})
  = \tilde{A} \frac{J^{\prime}}{J_{\bot}}
  + \tilde{B} (\frac{J^{\prime}}{J_{\bot}})^2.
\end{equation}
and $\tilde{A}$ and $\tilde{B}$ temperature independent constants.
The high temperature expansion again fixes
$\tilde{A}=2$, and fits to the QMC susceptibility data
in the temperature range $1.0 \leq T/J_{\bot} \leq 1.5$
give 
\begin{equation}
  \tilde{B} = 0.6940(3).
\end{equation}

\begin{fullfigure}
  \begin{center}
    
    \psbox{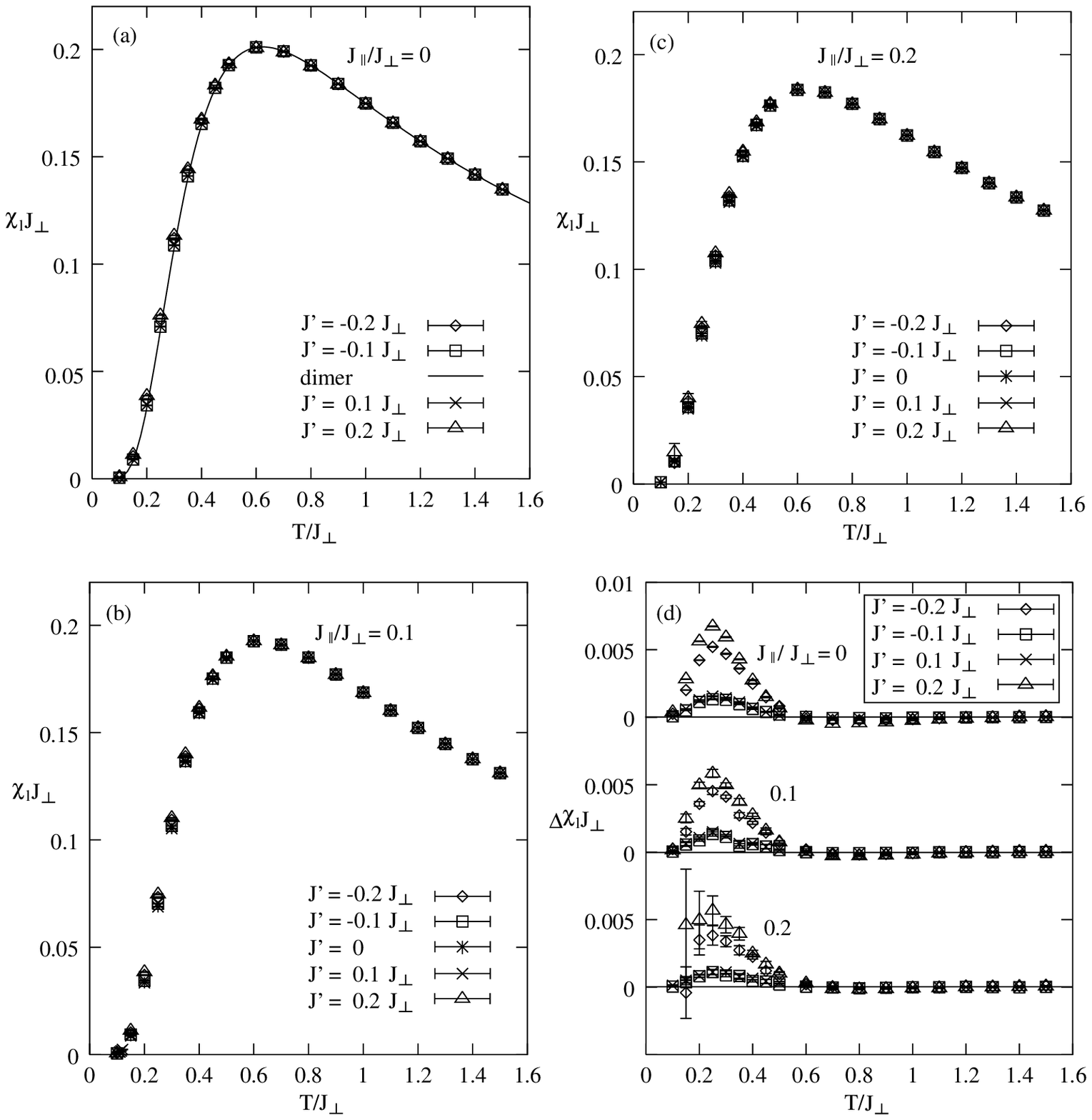}    
  \end{center}
    \caption{Scaling plots for the weakly coupled dimer regime for 
    couplings
      (a) $J_{\|}/J_{\bot} = 0$,
      (b) $J_{\|}/J_{\bot} = 0.1$ and (c) $J_{\|}/J_{\bot} = 0.2$.
      (d) shows the differences between the scaling curves}
    \label{fig:MFTScheck_dimer}
\end{fullfigure}
Scaling plots (Figure ~\ref{fig:MFTScheck_dimer}) show that the
agreement is not as good as in the coupled ladder case. This is not
very surprising, since, as can be seen in eq.(~\ref{eq:spingap_first})
and eq.(~\ref{eq:spingapgeneral}), the spin gap of weakly coupled
dimers depends more on the inter-dimer couplings than that of weakly
coupled ladders. Still the MFTS ansatz gives a reasonable analytic
parameterization of the susceptibilities of weakly coupled dimers.

\section{Comparison with Experiments}

In this section we wish to briefly compare our QMC results with
experimental measurements of the susceptibility of SrCu$_2$O$_3$ and
CaV$_2$O$_5$. A forthcoming publication\cite{Johnston2} will present
detailed fits and comparisons.

\subsection{{\rm SrCu$_2$O$_3$}}
With regard to fits of the experimental susceptibility measurements
on SrCu$_2$O$_3$ our main result is
that the inclusion of inter-ladder couplings does
not modify the previous estimate $J_{\bot}/J_{\|} \approx
0.5$\cite{Johnston} considerably. We cannot determine the value of
$J'$ from these fits since in the experimentally accessible
temperature range $T<650$K the dependence of the susceptibility on
$J'$ is weak. 

\subsection{{\rm CaV$_2$O$_5$}}
It was proposed previously that the magnetic susceptibility of
CaV$_2$O$_5$ can be fit by that of dimers.\cite{Onoda} We have
performed a series of fits on new experimental data of Isobe and
Ueda.\cite{Isobe}

One problem is that the samples are not pure CaV$_2$O$_5$ but contain
a few percent CaV$_3$O$_7$. By X-ray structural analysis Isobe and
Ueda determined that the current sample contains 4.1\%
CaV$_3$O$_7$. By subtracting the separately measured susceptibility of
CaV$_3$O$_7$\cite{Isobe} we obtained experimental data for pure
CaV$_2$O$_5$.

The main result of our fits is that the inter-dimer couplings $J_{\|}$
and $J'$ are both much smaller than $J_\bot$, and we can give some
constraints.

To determine the spin gap $\Delta$ we fitted the low temperature
experimental data to the expression for a
spin ladder at $T\ll\Delta$\cite{Troyer}
\begin{equation}
\chi(T) = \frac{C}{T-\Theta}+\chi_{0}+\frac{a}{\sqrt{T}}\exp(-\Delta/T),
\end{equation}
where the $g$-factor of Vanadium was determined to be $g=1.96$ by ESR
measurements.~\cite{Onoda,Isobe} The fit parameters were $C = 1.6
\times 10^{-3}$ cm$^3$K/mol V, $\Theta=-8$K and $\chi_0=1.23\times 10^{-5}$
cm$^3$/mol V. We use this fit to subtract the first term, which is a
Curie-Weiss term due to paramagnetic impurities, and the second,
temperature independent, constant contribution.

Next we fit the low temperature data ($T<200$K) to QMC results for an
isolated ladder for $J_{\|}/J_{\bot} = \pm 0.1$, $\pm 0.2$ and $\pm
0.3$. This fit is reasonable because we will see later, that we are in
the ladder case $|J'| \ll J_{\|}$, where the influence of $J'$ on the
low temperature susceptibility is negligible. In all cases we found
$J_{\bot}\approx670$K.
\begin{figure}
  \begin{center}
    \psbox{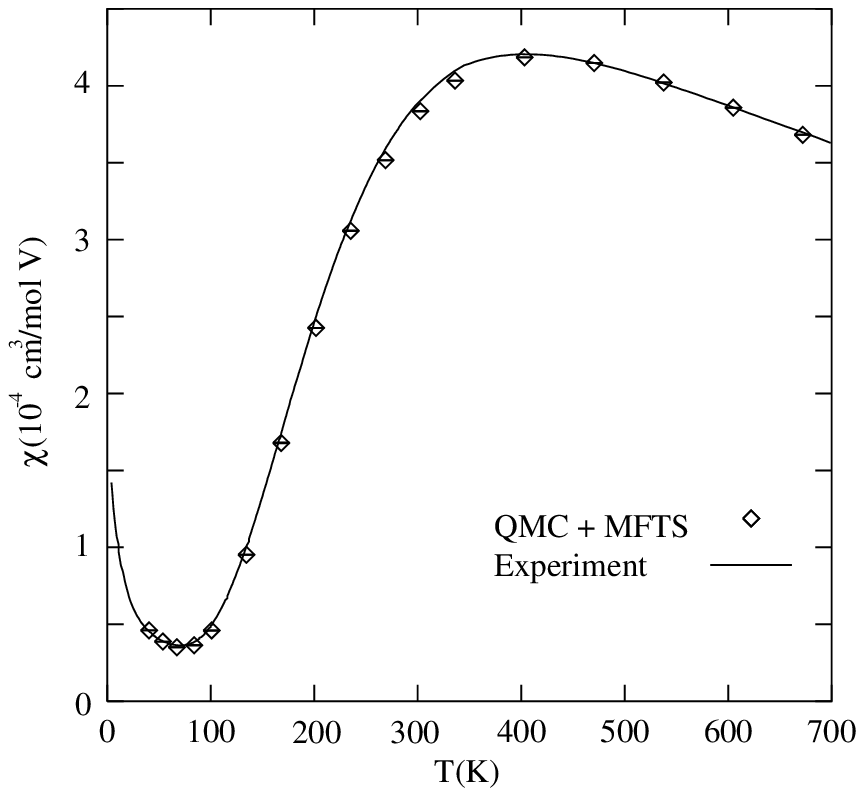}
  \end{center}
  \caption{Fit of the uniform magnetic susceptibility
  for CaV$_2$O$_5$, assuming $J_{\|}/J_{\bot}=0.1$. The fit parameters
  are $J_{\bot} = 672$K, $J_{\|} = 67$K and  $J^{\prime} = 45$K}
  \label{fig:CaV2O5}
\end{figure}

Next, for each of the above ratios of $J_{\|}/J_{\bot}$ with
$J^\prime/J_{\bot} = 0$ we use the 
MFTS eq. (\ref{eq:MFTS_dimer}) to fit the susceptibility in the range
$200{\rm K}<T<700{\rm K}$ and determine $J'$. Good fits are obtained
only for $J_{\|}/J_{\bot}=0.1$ and $0.2$, but not for
$J_{\|}/J_{\bot}\le0$ or $J_{\|}/J_{\bot}=0.3$. Assuming
$J_{\|}/J_{\bot}=0.1$ we obtained $J_{\bot}\approx670$K,
$J_{\|}\approx67$K and $J'\approx 45$K. This fit is shown in
Fig. \ref{fig:CaV2O5}. For the other ratio $J_{\|}/J_{\bot}=0.2$ we
got an equally good fit with $J_{\bot}\approx665$K,
$J_{\|}\approx135$K and a ferromagnetic $J'\approx -25$K.

Given the good quality of both fits it is hard to determine the exact
values of the couplings $J_{\|}$ and $J'$ from fits to the uniform
susceptibility alone. But we can give some estimates and constraints:
$J_{\bot}\approx 670$K, $0{\rm K}<J_{\|}<200$K and $J'+J_{\|}\approx
110$K.  The dispersion relation of the magnons however depends
sensitively on the ratio $J_{\|}/J'$ and could offer a way to
determine these couplings more precisely.

\section{Conclusions}

We have studied the trellis lattice Heisenberg model by quantum Monte
Carlo and a perturbation expansion around the dimer limit. We
confirmed that for weakly coupled ladders the influence of the
frustrated inter-ladder coupling on the spin gap is small.

We calculated the uniform susceptibility by QMC simulations
and found that a mean field-type scaling ansatz gives a reasonable
analytic parameterization of its temperature dependence in the whole 
temperature range.

This parameterization was in turn be used to fit experimental
measurements on the compounds SrCu$_2$O$_3$ and CaV$_2$O$_5$. For
SrCu$_2$O$_3$ it was found that the inter-ladder coupling does not
significantly modify the previous estimate of the intra-ladder
coupling ratio $J_{\bot}/J_{\|} \approx 0.5$.\cite{Johnston} CaV$_2$O$_5$
was confirmed to be a weakly coupled dimer system with inter-dimer
couplings about an order of magnitude smaller than the intra-dimer
coupling.

\section*{Acknowledgements}
The authors would like to thank M. Isobe and Y. Ueda for providing us with
new and higher quality measurement data on  CaV$_2$O$_5$ and CaV$_3$O$_7$.
The QMC program was written in C++ using a parallelizing Monte Carlo
library developed by one of the 
authors.\cite{alea}
The calculations were performed on the HITACHI SR2201 massively parallel 
computer of the University of Tokyo and of the Center for
Promotion of Computational Science and Engineering of Japan Atomic
Energy Research Institute.
Ames Laboratory is operated for the U.S. Department of Energy
by Iowa State University under Contract No.\ W-7405-Eng-82.
The work at Ames was supported by the Director for Energy
Research, Office of Basic Energy Sciences.


\begin{thebibliography}{99}
\bibitem{Daggot-Rice}
  E. Dagotto and T.M. Rice,
  Science {\bf 271} (1996) 618.
\bibitem{M.Azuma}
  M. Azuma, Z. Hiroi, M. Takano, K. Ishida and Y. Kitaoka,
  Phys. Rev. Lett. {\bf 73} (1994) 3463.
\bibitem{Nature}
  Z. Hiroi and M. Takano, Nature {\bf 377}, 41 (1995).
\bibitem{H.Iwase}
  H. Iwase, M. Isobe, Y. Ueda and H.Yasuoka,
  J. Phys. Soc. Jpn. {\bf 65} (1996) 2397.
\bibitem{142441}
  M. Uehara {\it et al.}, J. Phys. Soc. Jpn. {\bf 65}. 2764 (1996).
\bibitem{lacuo}
  B. Normand and T.M. Rice, Phys. Rev. B {\bf 54}, 7180 (1996);
  M. Troyer, M.E. Zhitomirsky and K. Ueda, Phys. Rev. B {\bf 55}, R6117
  (1997).
\bibitem{Johnston}
  D. C. Johnston,
  Phys. Rev. B {\bf 54} (1996) 13009.
\bibitem{Johnston2}
  D.C. Johnston {\it et al.}, in preparation.
\bibitem{Onoda}
  M. Onoda and N. Nishiguchi,
  J. Solid State Chem. {\bf 127} (1996) 359.
\bibitem{barnes92}
  T. Barnes, E. Dagotto, J. Riera and E.S. Swanson,
  Phys. Rev. B {\bf 47}, 3196 (1993).
\bibitem{reigrotzki}
  M. Reigrotzki, H. Tsunetsugu and T.M. Rice, J. Phys. Cond. Matt. {\bf 6}, 9235 (1994).
\bibitem{Troyer}
  M. Troyer, H. Tsunetsugu and D. W\"{u}rtz,
  Phys. Rev. B {\bf 50}, (1994) 13515.
\bibitem{Lanczos}
  C. Lanczos, J. Res. Natl. Bur. Stand. {\bf 45}, 225 (1950).
\bibitem{Evertz}
  H.G. Evertz, G. Lana and M. Marcu,
  Phys. rev. Lett. {\bf 70}, (1993) 875.
\bibitem{Evertz2}
  H.G. Evertz,
  cond-mat/9707221,
  to be published in {\it ``Numerical Methods for Lattice
    Quantum Many-Body Problems''},
  ed. D.J. Scalapino, Addison Wesley Longman, Frontiers in Physics.
\bibitem{Ammon}
  B. Ammon, H.G. Evertz, N. Kawashima, M. Troyer and B. Frischmuth,
  cond-mat/9711022.
\bibitem{Barnes2} T. Barnes and J. Riera, Phys. Rev. B {\bf 50}, 6817
  (1994).
\bibitem{Isobe}
  M. Isobe and Y. Ueda, unpublished data.
\bibitem{alea}
  M. Troyer, B. Ammon and E. Heeb, {\it Parallel object 
    oriented Monte Carlo simulations}, preprint.

\end{thebibliography}
\end{document}